\documentclass{aa}
\usepackage{graphics}
\input psfig.sty

\begin{document}


\title{On the origin of X-shaped radio-sources:\\
new insights from the properties of their host galaxies
\thanks{Based on observations obtained at
 the Italian Telescopio
Nazionale Galileo (TNG) operated on the island of La Palma by the Centro
Galileo Galilei of the CNAA (Consorzio Nazionale per l'Astronomia e
l'Astrofisica) at the Spanish Observatorio del
Roque de los Muchachos of the Instituto de Astrofisica de Canarias.}}

\author{
A. Capetti \inst{1} \and S. Zamfir \inst{1,2}
P. Rossi \inst{1} \and G. Bodo \inst{1} \and C. Zanni\inst{3} \and
S. Massaglia \inst{3}}

\institute{Osservatorio Astronomico di Torino, Strada Osservatorio 20, 
I-10025 Pino Torinese, Italy 
\and
Universitatea din Bucuresti, Facultatea de fizica atomica si nucleara, 
Bucuresti-Magurele, PO-BOX MG-11, Romania
\and
Istituto di Fisica Generale, Universit\'a di Torino, Via Giuria 1, 
10125 Torino, Italy}

\offprints{capetti@to.astro.it}

\date{Received ...; accepted ...}

\titlerunning{On the origin of X-shaped radio-sources}
\authorrunning{Capetti et al.}  

\abstract{A significant  fraction of extended radio  sources presents a
peculiar  X-shaped radio  morphology: in addition  to the  classical
double lobed structure, radio emission is also observed along a second axis
of symmetry in the form of  diffuse wings or tails.  We
re-examine  the   origin  of  these  extensions   relating  the  radio
morphology to the properties  of their host  galaxies.  The
orientation of the wings shows  a striking connection with the
structure of the  host galaxy as they are  preferentially aligned with
its minor axis. Furthermore, wings  are only observed in galaxies
of high projected ellipticity. 
Hydrodynamical simulations of the radio-source 
evolution show that X-shaped radio-sources naturally form 
in this geometrical situation: as 
a jet propagates in a non-spherical gas distribution, 
the cocoon surrounding the radio-jets 
expands laterally at a high rate producing wings of radio emission, in 
a way that is reminiscent of the twin-exhaust
model for radio-sources.
}

\maketitle

\section{Introduction}
\label{intro}

Extended radio sources have  been historically classified on the basis
of their radio morphology, the main division being based on their edge
darkened or edge brightened  structure that leads to the identification
of the Fanaroff-Riley classes I  and II (Fanaroff \& Riley 1974).  The
characteristic  structure of  FR~II sources  is dominated  by  two hot
spots located  at the edges  of the radio  lobes that, in  most cases,
show  bridges of  emission linking  the core  to the  hot  spots.  The
presence  of significant  distortions  in the  bridges was  recognized
since early interferometric  imaging of 3C sources (see  e.g. Leahy \&
Williams 1984). Distortion  in FR~II can be classified  in two general
classes: mirror  symmetric ( or  C-shaped) when the bridges  bend away
from the galaxy  in the same direction, or  centro symmetric when they
bend  in opposite  directions forming  an X-shaped  or  Z-shaped radio
source, depending  on the  location of the  point of insertion  of the
wings.   In  many  X-shaped  sources  the  radio  emission  along  the
secondary axis, although more  diffuse, is still quite well collimated
and can be even more extended than the main double lobed structure.

C-shaped  morphologies  are   observed  also  in  FR~I  radio-galaxies
although  in these sources  the distortions  affect their  jets rather
than their  lobes and they  give rise to  the typical shape  of Narrow
Angle  Tails, where  the opposite  jets bend  dramatically  and become
almost parallel one  to the other. Another common  morphology for FR~I
is that  of centro-symmetric S-shaped sources.   Conversely, there are
no examples of X-shapes among FR~I.

There is now a general  agreement that the C-shaped radio sources form
when they are  in motion with respect to the  external medium: jets or
bridges are bent  by the ram pressure of  the surrounding gas.  Models
successfully reproduced the morphology of FR~I Narrow Angle Tails (see
e.g. O'Dea \&  Owen 1986) and the extension of  this scenario to FR~II
bridges appears quite natural.

Concerning  the X-  or Z-shaped  sources several  mechanism  have been
proposed  for their  origin. Ekers  et al.  (1978) suggested  that the
tails of  radio emission  in one  of these sources,  NGC 326,  are the
result of the trail caused by  a secular jet precession (see also Rees
1978).  A similar model accounts for the morphology of 4C 32.25 (Klein
et al.  1995). In a  similar line, Wirth  et al.  (1982) noted  that a
change in the jet direction can be caused by gravitational interaction
with a  companion galaxy. Recently Dennett-Thorpe et  al. (2002), from
the analysis of spectral variations along the lobes, proposed that the
jet reorientation  occurs over short time  scales, a few  Myr, and are
possibly associated to instabilities  in the accretion disk that cause
a rapid  change in the  jet axis. In  all these models,  the secondary
axis of radio emission represents a  relic of the past activity of the
radio source.

An  alternative  interpretation was  suggested  by  Leahy \&  Williams
(1984) and  Worrall et  al. (1995).  They  emphasize the  role of  the
external  medium in  shaping radio  sources, suggesting  that buoyancy
forces can bend the back-flowing  material away from the jet axis into
the direction of decreasing external gas pressure.

In this Paper we present a different scenario based on evidence for a
strong  connection between the  properties of  the radio  emission and
those of  the host galaxy, a  comparison that has been overlooked in
the  past but  that provides  crucial new  insights on  the  origin of
X-shaped radio-sources.   In fact, in Sect. \ref{host}  we compare the
radio and  host galaxy orientation  of X-shaped radio  sources showing
that there is a close alignment between the radio wings and the galaxy
minor axis.  We  also show that X-shaped sources  occur exclusively in
galaxies  of high  ellipticity.  Numerical  simulations,  presented in
Sect.  \ref{simulations}, provide  support to  the idea  that X-shaped
radio-sources form  naturally in this  geometrical situation. Finally,
in  Sect. \ref{discussion}  we  discuss the  implications of  our
results that are summarized in Sect. \ref{summary}.

\begin{table*}
\label{tab1} 
\caption{Optical and radio P.A. of the winged radiosources}
\hspace{1.5cm} 
\scriptsize
\begin{flushleft}
\begin{tabular}{l c c c c} \hline\noalign{\smallskip}
Name &  Optical P.A. & Wings P.A. & Offset & Radio P.A. \\
\noalign{\smallskip}
\hline\noalign{\smallskip}
3C 52     &    55   &  -65  & 60  &  25  \\
3C 63     &    80   &  -45  & 55  &  30  \\
3C 136.1  &   -80   &   10  & 90  & -70  \\
3C 192    &   -85   &   60  & 55  & -55  \\
3C 223.1  &    40   &  -40  & 80  &  15  \\
3C 315    &    35   &  -45  & 80  &  10  \\
3C 403    &    35   &  -50  & 85  &  85  \\
4C 12.03  &   -25   &   70  & 85  &  15  \\
4C 32.25  &    90   &   -5  & 85  &  60  \\
\noalign{\smallskip}
\hline
\end{tabular}
\end{flushleft}
\end{table*}

\section{Host galaxy properties of X-shaped radio sources}
\label{host}

\subsection{The sample}
\label{thesample}

We searched  in the literature  for X-shaped radio  sources.  X-shaped
distortion are quite  common in FR~II but we  prefer to concentrate on
the 9  objects in  which the secondary  radio axis is  well developed,
with a  size of at  least 80 \%  of that of  the main radio  axis.  We
therefore excluded objects (like 3C 274.1, 3C 341 or 3C 381, see Leahy
and  Williams  1984)  in which the radio  wings  are present but are
considerably  smaller.  The  selected sources  are listed  in  Table 1
where we present also the  relevant parameters of the radio source and
of the host galaxy.

The  properties  of the  host  galaxies  have  been derived  from  the
analysis  of  infrared  K  band  images  obtained  at  the  Telescopio
Nazionale Galileo in  July 2000 as part of  a complete infrared survey
of 3C radio-galaxies with redshift z  $<$ 0.3 (see Capetti et al. 2002
for details on  these observations). The data also  provide us with a
control sample  of 18  radio-galaxies (with the  further observational
constraint  of  being  located  in  the  sky  at  15$^h  <$  R.A.  $<$
02$^h$). 

\subsection{Host galaxy and radio wings orientation}
\label{orientation}

\begin{figure} 
\resizebox{\hsize}{!}{\includegraphics{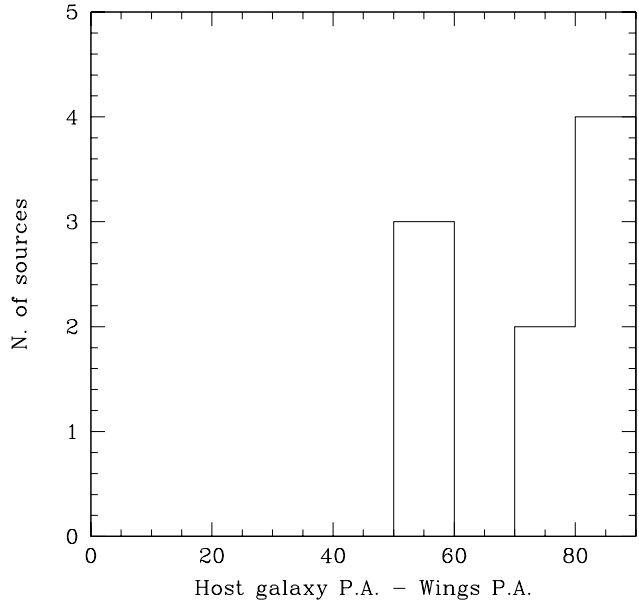}}
\caption{Relative orientation of the major axis of the host galaxy
and of the secondary axis of radio emission.}
\label{diff} 
\end{figure}

For each  source we determined the  orientation of the  radio wings by
connecting  the  outer  edges  of  the detected  radio  emission.   To
determine the  galaxy position  angle we used  the TNG K  band images.
For the  two winged radio  sources not belonging  to the 3C  sample we
instead  used optical  images  from the  Digitized  Sky Survey.   From
ellipse fitting of each galaxy we determined the behaviour of both the
position angle  and ellipticity of  the isophotes.  Typical  errors in
the  P.A. determination  are estimated  to be  $\sim  5^{\circ}$.  The
values  of radio  and host  galaxy  position angles  derived for  each
source are  given in Table  1 and a  histogram of their  difference is
shown  in Fig. \ref{diff}.  Clearly, the  PA difference  avoids values
smaller  than   $\Delta$  P.A.  $ <  55^{\circ}$  and   shows  a  strong
concentration at  high values with  an average value  of 75$^{\circ}$.
With  a  Kolmogorov-Smirnov  test   we  estimated  that  the  observed
distribution has only a probability of  5 \% to be drawn from a random
(uniform) distribution.

\begin{figure*} 
\resizebox{\hsize}{!}{\includegraphics{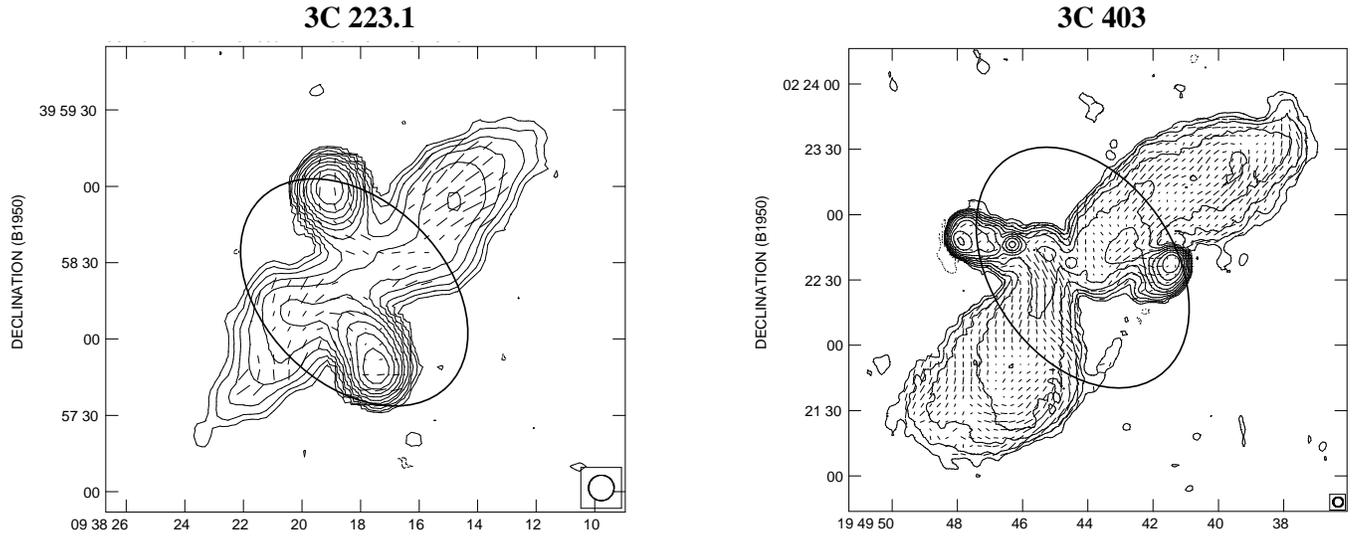}}
\caption{Superposition of the host galaxy shape (not in scale)
onto the radio maps for two X-shaped radio-sources.}
\label{super} 
\end{figure*}

Radio wings therefore appear to be oriented along a direction that
is almost  perpendicular to the  galaxy major axis  and consequently
along its minor axis (see Fig. \ref{super} for two representative examples). 
The  connection between radio and galaxy axis
becomes  stronger considering  the  effects of  projection.  In  fact,
while for  a highly flattened galaxy the major
axis is always orthogonal to the perpendicular to the galaxy disk, for a more
complicated (and realistic) galaxy shape, such as a triaxal ellipsoid,
projection  can only  decrease  the  angle between  the  two axis  and
populate also  relatively lower values of $\Delta$  P.A., similarly to
what  is observed  in  our  distribution. We  then  conclude that  the
orientation  of the  radio  wings  and galaxy  major  axis is  closely
connected  and,  considering  the  effects  of  projection,  they  are
plausibly intrinsically perpendicular one with respect to the other.

\subsection{Radio morphology and host ellipticity}
\label{ellipticity}

\begin{figure} 
\resizebox{\hsize}{!}{\includegraphics{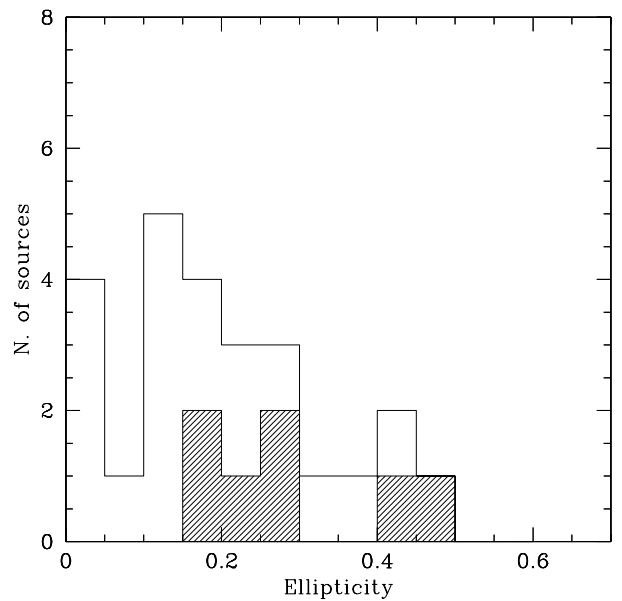}}
\caption{Host galaxy ellipticity of the X-shaped radio-sources 
(shaded histogram) compared
to a reference sample of 3C FR~II radio-galaxies. }
\label{elli} 
\end{figure}

A second result obtained from the analysis of the host galaxy images
is that winged radio sources appear to be associated to galaxies of high
ellipticity isophotes. To support quantitatively this visual impression in 
Fig. \ref{elli} we compare the distribution of ellipticity of the
seven winged source from the 3C (the DSS images of 4C 12.03 and 4C 32.25
do not allow us an accurate determination of ellipticity) with the control
sample. The ellipticity has been measured at a fixed distance
(3 \arcsec) from the galaxy center.
Clearly, the winged sources are only found for high values of ellipticity.
A Student test indicates that the probability that the ellipticity 
distribution of winged sources and the control sample are drawn from the same
population is only 2.8 \%.

\section{Numerical simulations} 
\label{simulations}

In the previous Section we showed the existence of a connection
between the host galaxy and radio properties: X-shaped radio sources
have their wings aligned with the galaxy minor axis and wings are
present only in galaxies of high ellipticity.  These results suggest
that the shape of the galaxy potential, which determines the
distribution of external gas, plays an important role in the evolution
of the radio-source.  To further investigate this issue we performed
numerical simulations of the propagation of a jet in a stratified
medium with an elliptical symmetry. Our aim is to show that an effect
exists and that it can originate morphologies analogous to the observed
X-shaped radio sources. Therefore we did not perform an extensive
parameter study but we considered two exemplificative cases, with the
jet oriented along and perpendicularly to the galaxy's major axis.

\begin{figure} 
\resizebox{\hsize}{!}{\includegraphics{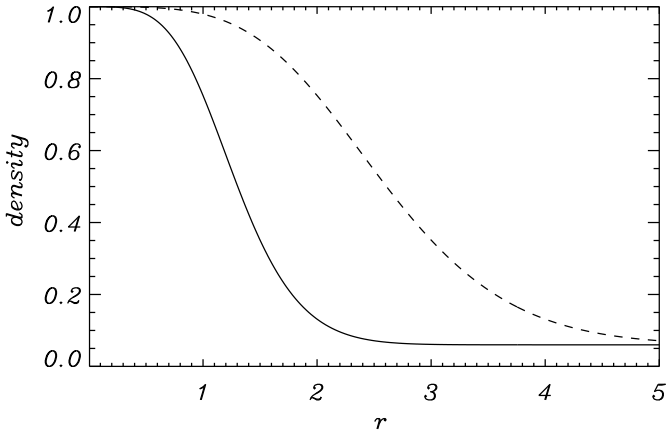}}
\caption{Equilibrium density profile of the adopted external gas distribution. 
The gas has an ellipsoidal shape and here we represent the density profiles 
along the major (dashed line) and minor (solid line) axis.}
\label{densprof} 
\end{figure}

The adopted external gas distribution has an ellipsoidal shape and, in
Fig. \ref{densprof}, we represent the density profiles along the major
and minor axis.  The width $r_c$ of the central plateau along the
major axis is chosen as our unit of length. The gas is isothermal and
the density distribution is kept in equilibrium by an appropriated
gravitational potential.

The integration is performed in
cylindrical geometry, the computational domain extends up to 6 $r_c$
 in the horizontal direction and up to 9 $r_c$ in the
vertical direction and is covered by $ 704 \times 1024 $ grid points.
We used symmetry boundary conditions at the bottom and left boundaries
and free outflow conditions at the top and right boundaries. The basic
equations of conservation have been integrated with a two-dimensional
version of the Piecewise Parabolic Method (Colella \& Woodward 1984).
The jet has a Mach number $M = 60$, defined with respect to the sound
speed in the ambient medium, and a density $3 \cdot 10^{-3}$ times the central
value in the external medium.

\begin{figure*} 
\resizebox{\hsize}{!}{\includegraphics{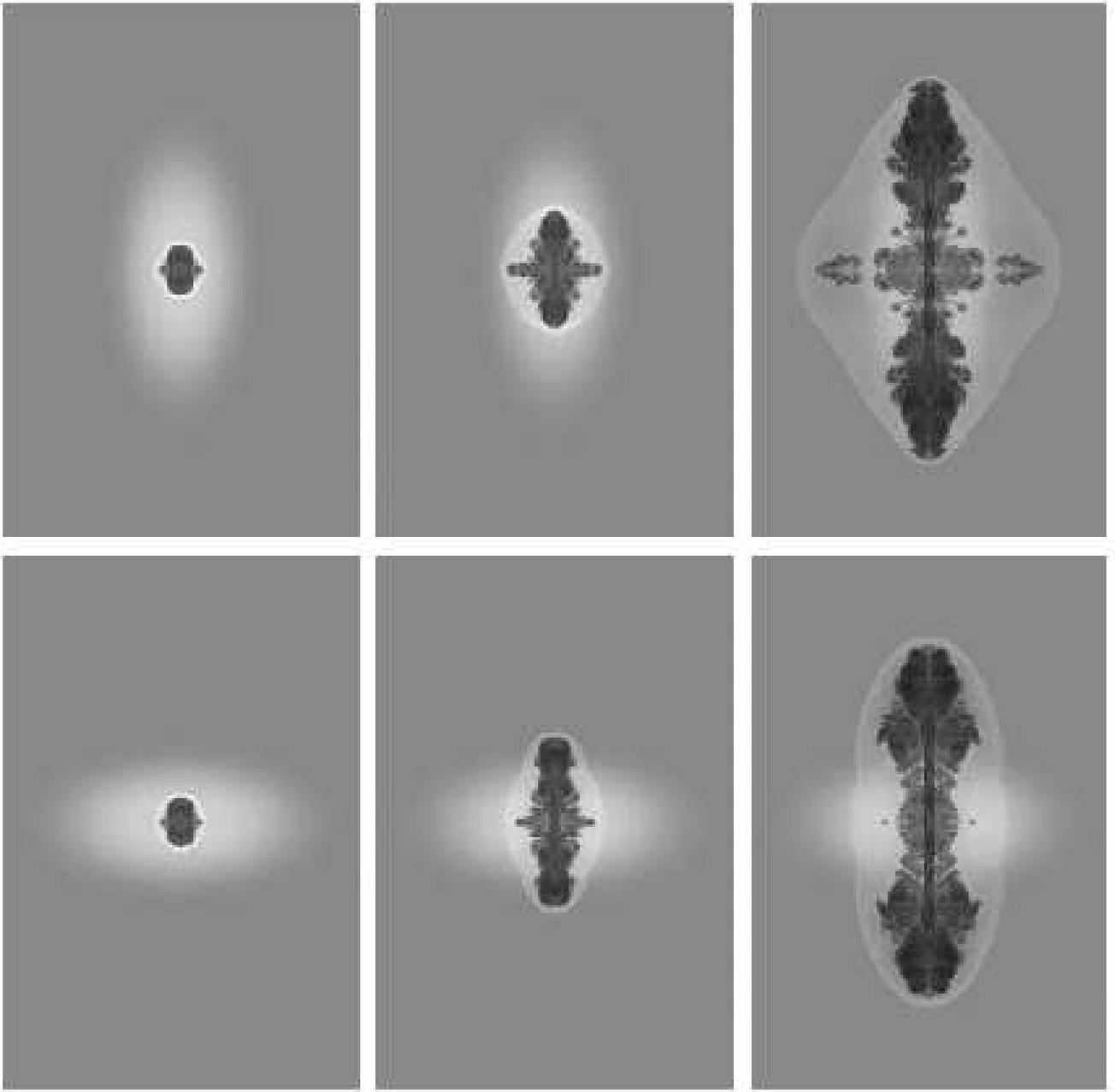}}
\caption{Results of numerical simulations: 
we show density images representing the evolution of
the radio-source in the stratified medium. The top
panels refer to the case when the jet is oriented parallel to the
major axis of the gas distribution, while the lower panels to the case
when the jet is oriented parallel to the minor axis.}
\label{sim} 
\end{figure*}

In Fig. \ref{sim} we show density images representing the evolution of
the radio-source in the anisotropically stratified medium.  The
computation has been performed on one quarter of the panels, but, for
clarity, we have reproduced also the symmetric regions.  The top
panels refer to the case when the jet is oriented parallel to the
major axis of the distribution, while the lower panels to the case
when the jet is oriented parallel to the minor axis.  

In the very
initial stages, when the radio source size is smaller than the central
density plateau of the external gas, the jet inflates a cocoon whose
shape is essentially spherical.  We can expect the gas to tend to
escape along the direction of the minor axis of the density
distribution. In the two cases examined, 
the difference of course is that we have the jet pushing
along a defined direction which can be parallel or perpendicular to
the minor axis. Therefore, at later times, we can expect different
morphologies. Indeed, when the jet is oriented along the
galaxy's major axis, the cocoon expands also laterally at relatively large
speed due to the faster decrease in the external density in this
direction. Furthermore, the distribution
of the external gas causes a collimation of the outflow. This collimation is
maintained also in the later stages of the evolution and produces an
ordered outflow along the galaxy's minor axis and the overall
morphology of the radio source is typical of the X-shaped sources.
Conversely, when the jet is parallel the minor axis, a classical
double radio source forms.  

We conclude that the development of wings
arises only when the jet is aligned with the major axis of the galaxy,
as in this situation the lateral expansion of the radio source is
favoured over the longitudinal expansion.

\begin{figure} 
\resizebox{\hsize}{!}{\includegraphics{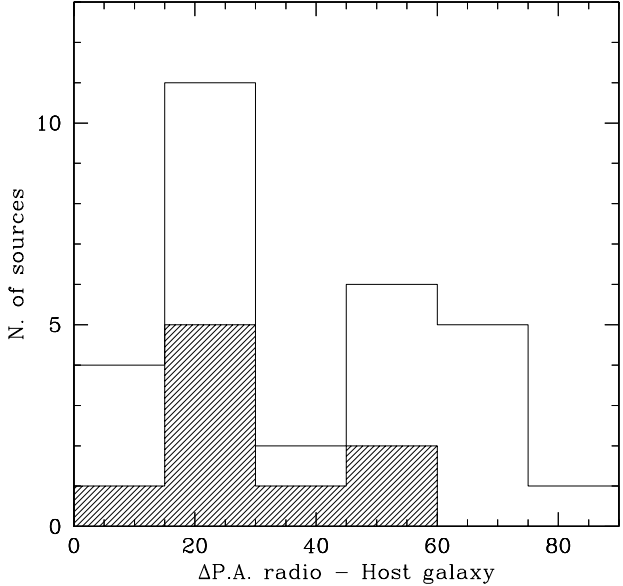}}
\caption{Relative orientation of the major axis of the host galaxy
and of the primary axis of radio emission.}
\label{major} 
\end{figure}

This result finds immediate confirmation in the observations:
winged radio-source show a clear alignment between the
principal radio and host axis
(see Fig. \ref{major}), an effect which is not shared by the
control sample. We conclude that wings naturally form in highly elliptical
galaxies when the major radio axis form a small angle with the host galaxy's
major axis.

\section{Discussion} 
\label{discussion}

The results presented in the previous sections show that 
the origin of X-shaped radio-galaxies is a natural consequence of
the expansion of the cocoon formed by the radio-source
in a stratified medium with an elliptical
distribution. This situation is reminiscent of the
scenario envisaged by Blandford and Rees (1974) in their twin-exhaust
model: we have a bubble of hot and light gas inside an elliptical
distribution of density and pressure. 
The bubble expands laterally (with respect to the jet axis)
at a high rate due to the faster decrease in the external density in this
direction with respect to the jet's direction. 
The external gas plays a second crucial role as it also collimates the 
cocoon expansion forcing the growth of well defined radio wings.
Note that actually in our simulation, given the adopted cylindrical
symmetry, a circular equatorial outflow is produced, that suffers a
strong dilution. However, in a three-dimensional case, 
it is likely that the gas will escape in one dimension only, 
along the galaxy's minor axis, therefore maintaining a stronger thrust
with respect to the outflow appearing in our results. 
For this reason we expect the cocoon lateral expansion
to be even more prominent that suggested by the simulations.

The wings expansion progresses until the jet exits the central region 
of high density. At this stage the head of the radio-source advances
at a substantially higher speed than in the earlier phases; the material 
backflowing towards the nucleus is not able to reach back the central cavity
and stops to provide pressure support for the lateral expansion.  
Clearly when and how this will affect the radio-source morphology
can only be assessed with more detailed three-dimensional simulations
and with better information on the external gas distribution.

It is important to stress that the physical mechanism that we propose as 
the origin of X-shaped radio-sources
is different from those suggested in the past,
not only from those suggesting jet reorientation discussed below, 
but also from those 
in which the wings are formed by other hydrodinamical effects.
In particular Worrall et al. (1995) and Leahy \& Williams (1984) 
emphasized the importance of buoyancy forces acting on the backflow.
While buoyancy can also be important in our framework, we suggest that
the driving mechanism at the origin of the radio-wings is the presence
of an overpressured cocoon. In fact, as noted above, it might
be problematic to channel the backflowing plasma in the wings 
once the radio-source is fully developed, as these models predict.

The results presented here, i.e. the connection between  
the properties of the host galaxy and the presence of a 
X-shaped radio morphology, 
appear to represent a significant problem for the competing model
based on jet re-orientation. In fact it remains to be explained why
the re-orientation should occur in a small fraction
of radio-galaxies and preferentially in objects 
in which the jet is aligned with the galaxy minor axis.
A second observational evidence, which
is difficult to account in the framework of jet realignment, is 
the lack of winged FR~I. 
In our scenario this is readily interpreted as FR~I do not form the
overpressured cocoons that are one of the necessary ingredient to lead to
the formation of the wings.

Constraints on the wings origin may come from the
radio spectral index mapping of these radio-sources as this 
can be related
to the age of the different radio-structures. 
In the context of jet re-orientation, 
wings represent aged relic of previous activity and 
the similarity of
the spectral index on the main lobes and on the wings 
(Dennett-Thorpe et al. 2002) can be used
to set limits on the realignment timescale.
Conversely, in our scenario the radio wings are formed by jet material that
flowed back to the central regions of the radio source and 
is then channeled away by the pressure excess of the cocoon. 
During this process the relativistic plasma experiences 
radiative losses 
(via synchrotron radiation and up-scattering of cosmic background photons)
but, on the other hand, wings are
continuously replenished with newly injected
plasma and are tracers of an active outflow. In this situation
significant reacceleration can occur due to
shocks or turbulence and no simple prediction on the spectral index behaviour
can be obtained.

Our results can also be tested by looking for counter examples, i.e. 
a source with the proper geometrical setup but without wings. For example,  
we found that wings are only present in galaxies with high ellipticity
($\epsilon  > 0.17$).   However, there  are 6  objects in  the control
sample with isophotes ellipticity larger than this values, but without
radio wings. It is worth discussing these objects individually. In two
cases (3C 319 and 3C 456)  the jets are almost aligned with the galaxy
minor  axis (within  20$^\circ$ and  30$^\circ$ respectively)  and, as
shown above,  in this geometrical  configuration we do not  expect the
formation of  wings; 3C 357  shows a centro-symmetric  lobe distortion
(see Harvanek  \& Hardcastle 1998)  although not well developed  as in
the X-shaped sources; 3C 326 is  a giant radio-galaxy 
(its size exceeds 2 Mpc, Willis \& Strom 1978) 
and as such it probably evolved to a stage in which it is not influenced
by the gas distribution on galactic scale;  
for the  remaining two  objects, 3C  327 and  3C 452,  there are
simply no high  quality images in the literature to be used to
explore the morphology  of their lobes. Therefore we  could not find
radiosources representing strong counter examples conflicting with our
model.

In our scenario two geometrical requirements must be fulfilled to produce
a winged radio-source.
This is particularly important when one wants to assess the relevance
of this class. In fact,
X-shaped morphology are observed  in a relatively small fraction (7\%)
of  FR~II radio-sources (Leahy \& Parma 1992). 
This  value suggests  that they  represent an
intriguing but somewhat limited class.  On the contrary, we argue that
they  develop essentially in  all situations  that fulfill  the proper
geometrical requirements:  i) jets  must form a  large angle  with the
galaxy minor  axis and ii) wings  can occur only in  galaxy of large
ellipticity.   Furthermore,  projection   can  hide  genuine  X-shaped
sources  when e.g. the  wings project  onto the  lobes, or  either the
wings or  the jets are aligned  with the line of  sight.  Finally, our
definition  of  X-shaped  radio-source  is quite  strict  and  several
additional objects (as e.g. those cited in Sect. 2) might qualify if a
looser definition  is adopted. Note  also that in several  cases radio
maps adequate  to show  the presence of  diffuse wings are  simply not
available. 
Thus we argue that X-shaped radio-sources should not be viewed as 
peculiar objects but as the natural manifestation of the growth of the
cocoon in a flattened distribution of external gas.

Finally, it appears that the proposed model can find application
also in interpreting the evolution of other classes of radiosources. 
For example, the rapid lateral expansion due to gas stratification
has been invoked by Capetti et al. (1999) to explain the dynamical 
properties of
emission line cocoon formed by the radio-jets in the Seyfert galaxy Mrk 3. 
Furthermore, a flattened distribution of external
gas can be produced by, e.g., the presence of 
a companion galaxy in a common envelope. 
In this case the external gas will have a symmetric distribution
on galactic scales and an elliptical shape only on larger scales.
Its asymmetry will start to affect the cocoon only 
when the radio-source has expanded to a size similar to the separation
of the galaxies forming the double system; 
this might explain the Z-shaped morphology of
the radio-source NGC 326 whose host is one component of a galaxies pair 
(see Murgia et al. 2001).

\section{Summary and conclusions}
\label{summary}

We presented evidence that the orientation of the secondary axis
of radio emission in X-shaped radio-sources is closely linked to
the orientation of the host galaxy. More specifically, considering the effects
of projection, the radio wings appear to be parallel to the 
galaxy minor axis. This relationship suggests a causal connection
between the galaxy property and the origin of the X-shaped radio morphology.
Such a connection is strengthened noting that X-shaped radio sources
are found only in galaxy of high ellipticity. 
We then  argue that X-shaped  radio-sources naturally form when  a jet
propagates  in a  non-spherical  gas distribution. In  this case  the
cocoon expansion along the direction of maximum pressure gradient (the
galaxy  minor axis) can  occur at  a similar  (or higher)  rate than
along the  radio source main  axis. 
The situation is reminiscent of the twin-exhaust
scenario where a bubble of hot and light gas inside an elliptical
distribution of density and pressure 
expands laterally at a high rate producing wings of radio emission.
Hydrodynamical simulations  of the radio-source  evolution in this 
situation support this scenario.

From the point of numerical simulations, 
a three dimensional analysis is clearly needed to investigate in more detail
the radio source behaviour, but it might be envisaged that with more 
realistic simulations it might be possible to constraints
the jet's properties based on the wings formation. 
In particular it will be important 
to perform  a thorough exploration of the jet's parameters space.
In fact, it is clear that the effects
of the asymmetric gas distribution will affect more jets of lower
kinetic power: at higher power the crossing time of the galaxy's
central regions will be shorter and this will in turn shorten the phase 
during which the cocoon can expand laterally. 
We thus expect that only jets in a limited region of the parameters 
space might produced winged radio-sources.

Finally, a very valuable information will come from mapping of the external gas
from high resolution X-ray images such as those that can now be obtained
with the new generation of X-ray satellites such as Chandra of Newton-XMM.

\begin{acknowledgements}
The authors acknowledge the Italian MIUR for financial support, grant
No. 2001-028773. The numerical
calculations have been performed at CINECA in BOlogna, Italy,
thanks to the support of INAF. This
research has made use of the NASA/IPAC Extragalactic Database (NED)
which is operated by the Jet Propulsion Laboratory, California
Institute of Technology, under contract with the National Aeronautics
and Space Administration. 
\end{acknowledgements}

\end{document}